# Waste forms for actinides: borosilicate glasses

Bernd Grambow

SUBATECH (Ecole des Mines de Nantes, Université de Nantes, CNRS-IN2P3), 4 rue Alfred Kastler, 44307 Nantes, France

**Introduction**

The fate of actinides in the nuclear fuel cycle comprises all steps from mining and enrichment of uranium to fuel fabrication and fuel use in nuclear reactors until the moment when no reuse is anymore foreseeable, the moment where some uranium becomes waste together with other actinides. The actinides in the waste are formed by neutron capture during reactor irradiation of the nuclear fuel. The largest fractions of actinide containing wastes are in the form of spent nuclear fuel, for which in many countries no reuse is foreseen. Spent fuel is then conditioned by packaging in safe long-term tight containers. In other countries with closed nuclear fuel cycles, reprocessing industries extract uranium and plutonium from the spent nuclear fuel for reuse in the form of mixed oxide fuel (MOX) in nuclear reactors. Extraction yields are always lower than 100% and traces of uranium and plutonium remain in the waste stream. This high level liquid reprocessing waste will become vitrified in order to obtain a stable borosilicate waste glass matrix that provides protection against environmental dispersion. Since vitrified waste is in a solid form, transportation, storage and final geological disposal are facilitated. The glass products contain the non-extracted fraction of uranium and plutonium together with the minor actinides Am, Np and Cm and fission products such as ($^{90}$Sr, $^{99}$Tc ou $^{137}$Cs) . Higher actinide loadings may occur in the future if reprocessing of MOX fuel is foreseen. Nuclear waste glasses were also proposed as host waste matrix to immobilize excess weapons plutonium in proliferation resistant way.

Considerable worldwide experience exists in the vitrification of high level waste. To fabricate a glass, typically, about 20% of waste will become melted at very high temperatures (>1000°C) with inactive glass forming ingredients ($SiO_2$, $Al_2O_3$, $B_2O_3$, $Na_2O$…) to form a homogeneous matrix phase in which the radionuclides becomes dissolved in solid solution and atomically bonded when the glass cools to a rigid solid. The glass fabrication process must be performed in shielded cells with very effective off gas cleaning to protect operators and environment. The molten glass is remotely poured into metal (stainless steel or copper) canisters in preparation for temporary storage, transportation, and ultimate disposal.

In the long term, the safest way to deal with actinide containing nuclear waste (vitrified waste or spent fuel) is geological disposal. Geological disposal comprises the removal of the waste from the biosphere by putting it in stable deep inaccessible host rock formations. The disposal locations are isolated from biosphere by engineered barriers while dense rock characteristics and low groundwater flow assure that any potential return of radionuclides to biosphere by groundwater transport will be retarded for many hundreds of thousands of years. If the actinides (and other radionuclides) are conditioned within a glass matrix and cast into metal canisters, actinides can only be released to the ground water if the canister and the glass matrix becomes dissolved or corroded and this alteration process will take many hundreds of thousands of years. Even if the glass finally becomes dissolved or corroded entirely in ground water, large quantities of actinides will in all likelihood remain immobile by being bound to secondary phases, which form as solid corrosion products on the glass surface during the alteration process. Typical glass corrosion products are gel like and crystalline phases such as silicates and oxy-hydroxides. Once released to groundwater, the fluid will carry away the actinides but the transport distances are very short due to sorption on mineral surfaces which are present along the transport path. To predict the performance of the actinide containing waste glass, one needs to understand the dissolution process of the glass matrix, the thermodynamic and kinetic constraints which lead to actinide fixation in secondary reaction products as well as the short and long term transport processes.

**Incorporation of actinides in the glass structure**

The incorporation of actinides in the glass has been studied by many authors. Like for other glass constituents, their incorporation in the glass matrix depends on electronegativity, on ionic radius and on the field strength of the uranium ion, hence also on the oxidation state of the metal ion. In general, the actinides become incorporated at the molecular level, exhibiting different coordination environments with oxygen atoms within the three dimensional glass network. The glass network is established by component oxides which play the role of so called network formers creating oxygen-bridges between small highly charged cations with small coordination number like $SiO_2$, $B_2O_3$, $P_2O_5$, while network modifiers ($Na_2O$, $Li_2O$, $CaO$, $La_2O_3$ etc.) transform the bridging oxygen (BO) bonds into non-bridging oxygen (NBO). Sometimes cations can play an intermediate role ($Al_2O_3$, $UO_2$, $ThO_2$..) in glass. By neutron scattering on borosilicate glasses Sinclair et al. (1983) have shown that the glass structure is hardly affected whether radioactive waste oxides are incorporated in the glass structure or not. Based on $^{29}$Si and $^{11}$B MAS NMR and IR studies it has been shown by Mishra et al. (2006) that the borosilicate network is not affected by a $ThO_2$ incorporation as large as 18.6 wt%.

To ensure optimum actinide confinement, the glass must be as homogeneous as possible. Excess loading of actinides beyond their solubility limit may lead to precipitation of actinide containing phases. The current actinide loading limit is for example 0.4 wt% for French nuclear waste glasses [Cachia et al. 2006] but strong interest exists to increase these limits. Lanthanide-Borosilicate glass is capable to dissolve 7 or 10% of $PuO_2$ in the glass structure [Bibler et al. 1996, Shiryaev et al 2009]. At higher Pu contents (15wt%) some $PuO_2$ remains undissolved in the glass structure. Britholite ($Ca_2Nd_8(SiO_4)_6O_2$, $Ca_2La_8(SiO_4)_6O_2$; based on ionic radii of $Nd^{3+}$, $La^{3+}$, and $Pu^{3+}$, an extensive range of solubility for $Pu^{3+}$ substitution for the Nd or La, particularly on the *6h* site, is expected) was also observed as a precipitated phase. In case of precipitation of $PuO_2$, and britholite there is Pu partitioning between glass phase and these crystalline phases [Shiryaev et al. 2009]. In the Nd-La system simulating trivalent actinide incorporation for glasses cooled from 1200ºC to room temperature at 350ºC·h$^{-1}$ the solubility limit is located above 12.5 wt.% and below 15 wt.% for $Nd_2O_3+La_2O_3$. It corresponds to the concentrations in the glass beyond which appearance of rare-earth apatite-like silicates $Ca_2RE_8(SiO_4)_6O_2$ (RE: rare-earth) is observed in the glass. Corresponding Am containing glasses lead to precipitation of Am rich apatite like crystals [Kidari et al. 2012].

The oxidation state of the actinides has a significant effect on their solubility in glasses as noted by Schreiber (1982a and b): for example, the U(VI) solubility is close to 40 wt% while that of U(IV) does not exceed 9 wt%. Cerium solubility in glasses was increased significantly by a reducing melting process from 0.95 for Ce(IV) to 13.00 wt% for Pu(III) [Cachia et al. 2006]. Similarly, Plutonium solubility in the glass can be increased to values above 4 wt% by reducing Pu(IV) in the glass melt to

Pu(III) [Deschanels et al. 2007]. Using electron energy loss spectroscopy (EELS), Fortner et al. (1997) found that the redox state of cerium doped to 7 wt% could be varied by a suitable choice of alkali elements in the glass formula. Plutonium-solubility could be enhanced in similar manner to values beyond 5wt%.

The details of phase separation of actinides in glasses are not fully understood as they involve phenomena such as glass melt cooling history, nucleation and nanoscale growth, potential liquid-liquid phase separation in the melt and redox (<u>red</u>uction/<u>ox</u>idation) interaction between other polyvalent glass constituents. During glass melting, the equilibrium between redox states of the multiple component oxides is complex and depends on the base glass composition as well as the melting environment. Cerium was found [Yang et al 2010] accommodated in both its valences within up to four different phases, including crystalline Ce(IV) dendrites and amorphous droplets of different size regime. The Ce valency, has a distinct dependency on the presence of other multi-valent elements at concentrations of several mol%, and on precipitation of crystals (whether direct or indirect via change of the residual glass composition).

If the glass is molten under oxidizing conditions, uranium exists in the glass mainly in its hexavalent state [Lam et al. 1979], while its formation under reducing conditions in graphite containers leads to tetravalent uranium in the glass [Antonini et al. 1984]. The establishment of reducing conditions in the melt process can be facilitated by adding sugar, or other reductants such as formic acid, glucolic acid, urea, oxalic acid or even metallic species to the glass melt [Veal et al. 1987, Jantzen 2013]. Nevertheless, when a glass is molten by adding $UO_2$, some U(IV) may remain retained in the final glass product. Ollier et al. (2003) analyzed such a glass (SON68) by XPS and photoluminescence and they found 80% U(VI) and 20% U(IV). The defense waste pilot facility (DWPF) in Savannah River runs at 20% Fe(II) and this results in a mix of U(IV), U(V) and U(VI) [Jantzen 2013]

The redox state of radionuclides in a polyvalent cation containing glass can also be modified by changing the temperature. For example, using Ce as surrogate for Pu (referring to ion radius and multi-valence), the temperature was observed to modify the Ce(III)/Ce(IV) ratio in the glass (68% Ce(III) at 1200°C and 88% Ce(III) at 1400 °C) [Jollivet et al. 2005]. Results of Lopez et al. (2003) show an increase in the solubility of the actinide surrogates (Ce, Hf, Nd) with an increase of glass melting temperature from 1200 to 1400°C. This effect was strongest for Ce and it is due to the reduction of Ce(IV) to the higher soluble Ce(III) following oxygen degassing at high temperature. Similarly, oxidation states of Fe and Mn becomes as well reduced [Jantzen 2013].

**Coordination environments of actinides in the glass**

*Trivalent actinides*: Europium (and probably also the trivalent actinides) appears in nuclear waste borosilicate glasses both in silicate and in borate environments [Ollier et al. 2003], Nd only on silicate sites [Jollivet et al. 2005]. The $Eu^{3+}$ cation is 6-coordinated with a Eu-O distance of the order of 2.31 Å, showing that it is a good surrogate of Cm in borosilicate glasses and $^{IV}B/^{III}B$ ratio decreases with europium oxide increase [Bouty et al. 2012]. A study of the coordination environment by optical absorption and EXAFS spectroscopies of Nd in Nd containing alumino-borosilicate glasses has shown [Quintas et al. 2008] that independent on the alkali/alkaline-earth ion types $Nd^{3+}$ sites are well-defined and similar in all studied glasses. The neodymium ion is surrounded by 7–8 non bridging oxygens (NBO), and the negative charge of the coordination shell is compensated by 2 to 3 alkalis, or

1 alkali and 1 alkaline-earth ions, per NBO. These data confirm that trivalent rare earth and probably actinide elements play the role of network modifiers in the glass structure.

*Tetravalent actinides:* The local first neighbor environment of Pu(IV) in the glass is nearly identical to $PuO_2$ the only difference being that the Pu-0 distance is slightly smaller in the glass [Knapp et al. 1984]. These authors reported 8+1 oxygen near-neighbor atoms. However, in 10wt% Pu containing Lanthanide-Borosilicate glass Pu(IV) was found 5 fold coordinated [Shiryaev et al. 2009]. In general, tetravalent uranium occurs in glasses in 6 to 8 fold coordination.

*Hexavalent actinides*: Uranium (VI) ions in glass occupies two distinct environments: typical uranyl and $UO_3$ type configurations [Ollier et al. 2003]. Jollivet et al. 2002 observed a $\gamma$-$UO_3$ environment both for bond distances and coordination number for two U(VI) sites in the glass SON68 and absence of U-U contributions to the spectra indicate that even for U contents as high as 3,4 wt% uranium remained homogeneously distributed in the glass.

Table 1 shows coordination environments observed by various authors and for various glasses including as well some data for glass surfaces altered in water.

Table 1: coordination environments of Actinides and rare earth elements in glasses and altered glass (gel) using EXAFS analyses

| Bond | Glass composition | Bond distance in A | Coordination number | Ref |
|---|---|---|---|---|
| Ce-O | Borosilicate glass SON68 (simplified) | 2.44 | 8 | Jollivet et al. 2005 |
| Nd-O | Borosilicate glass SON68 (simplified) | 2.48 | 8 | Jollivet et al. 2005 |
| Th-O | Borosilicate-Waste | 2..36 | 8..7 | Petit Marie et al. 1986 |
|  | Borosilicate |  |  | Petit Marie et al. 1986 |
|  | Monticellite (3%Th) | 2..34 | 7.8 | Farges et al. 1991 |
|  | Diopside (3%Th) |  |  | Farges et al. 1991 |
|  | Anorthite(3%Th) | 2.40 | 6.5 | Farges et al. 1991 |
|  | OliveneBasalt (3% |  |  | Farges et al. 1991 |
|  | TholeiteBasalt 3% | 2.41 | 6.3 | Farges et al. 1991 |
|  | Ryholite (3%Th) |  |  | Farges et al. 1991 |
|  | Albite (3%Th) | 2.37 | 5.0 | Farges et al. 1991 |
|  | Albite (1%Th) |  |  | Farges et al. 1991 |
|  |  | 2.35 | 6.0 |  |
|  |  | 2.35 | 5.9 |  |
|  |  | 2.36 | 6.6 |  |
|  |  | 2.36 | 6.5 |  |
|  |  | 2.33 | 5.2 |  |
| U(VI)-O | Na-Borosilicate | 2.3 | 5 | Graeves 1980 |
|  | Na trisilicate glass | 2.25 | 4 | Knapp et al. 1984 |
|  | $Na_2Si_3O_7$ | 2.25 | ? | Farges et al. 1991 |
| | $Na_{3,3}AlSi_7O_{17}$ | 2.25 | ? | Farges et al. 1991 |
| | SON68 bulk glass (site 1) | 2.2 | 4.0-4.2 | Jollivet et al.2002 |
| | SON68 bulk glass (site 2) | 2.31-2.32 | 4.0-4.2 | Jollivet et al.2002 |
| | SON68 gel surface (site 1) | 2.22-2.23 | 5.0-6.0 | Jollivet et al.2002 |
| | SON68 gel surface (site 2) | 2.38-2.41 | 5.6-6.6 | Jollivet et al.2002 |
| | GP WAK 1 bulk glass | 2.21 | 3.9 | Bosbach et al. 2009 |
| | GP WAK 1 gel surface (323 K leach) | 2.18 | 5.2 | Bosbach et al. 2009 |
| | GP WAK 1 gel surface (363 K leach) | 2.18 | 4.7 | Bosbach et al. 2009 |
| U(VI)=O | Na-Borosilicate | 1.9 | 2 | Greaves 1990 |
| | Na trisilicate | 1.85 | 2 | Knapp et al. 1984 |
| | $Na_2Si_3O_7$ | 1.78 | 2 | Farges et al. 1991 |
| | $Na_{3,3}AlSi_7O_{17}$ | 1.78 | 2.2 | Farges et al. 1991 |
| | Borosilicate waste glass | 1.75-1.80 | 2 | Petit-Marie et al. 1986 |
| | SON68 bulk glass (site 1) | 1.79-1.81 | 2.0-2.14 | Jollivet et al.2002 |
| | SON68 bulk glass (site 2) | 1.79-1.81 | 2.0-2.1 | Jollivet et al. 2002 |
| | SON68 gel surface (site 1) | 1.79-1.80 | 1.8-2.1 | Jollivet et al.2002 |
| | SON68 gel surface (site 2) | 1.79-1.80 | 1.8-2.1 | Jollivet et al. 2002 |
| | GP WAK 1 bulk glass | 1.77 | 1.8 | Bosbach et al. 2009 |
| | GP WAK 1 gel surface (323 K leach) | 1.81 | 2.1 | Bosbach et al. 2009 |
| | GP WAK 1 gel surface (363 K leach) | 1.79 | 2.0 | Bosbach et al. 2009 |
| U(V)-O | Albite | 2.19 | 4.7 | Farges et al. 1991 |
| | $Na_{3,3}AlSi_7O_{17}$ | 2.24 | 4.2 | Farges et al. 1991 |
| U(IV) | $Na_2Si_3O_7$ | 2.27 | 4.2 | Farges et al. 1991 |

**Radiation stability of actinide containing glasses**

Actinide containing glass is subject to strong irradiation effects during long term storage. Alpha decay of the actinides is responsible for most of the atom displacements. Spectroscopic studies (XAFS) of Pu-bearing glasses show an increase in disorder and bond lengths of oxide polyhedral [Hess et al. 1998] and stored energy [Weber et al. 1998] with increasing alpha decay dose.

External irradiation by 5MeV alpha particles produced by a cyclotron to doses of about $1.3 \cdot 10^{15}$ $\alpha/cm^2$ (corresponding to about $2 \cdot 10^{17}$ $\alpha \cdot g^{-1}$) showed no effect on glass leaching rates under silica saturated conditions and no defects were observed by positron annealing spectroscopy [Suzuki 2008]. The long-term behavior of actinide containing borosilicate glasses under a self-irradiation was estimated for integrated doses up to $4 \cdot 10^{18}$ $\alpha \cdot g^{-1}$. Varying the alpha dose rate in Pu doped glasses from $1.3 \cdot 10^{16}$ to $2.6 \cdot 10^{18}$ $\alpha \cdot g^{-1}$ by changing the $^{238}Pu/^{239}Pu$ ratio in the glass did not change the plutonium release rate or the glass dissolution rate measured by boron release [Wellman et al. 2005]. No significant effect was also observed by French teams on the initial alteration rate, swelling remained limited to about 0.5% and saturated for a dose of $2 \cdot 10^{18}$ $\alpha \cdot g^{-1}$. Borosilicate glass thus demonstrates a potential for the conditioning of plutonium [Deschanels et al. 2007]. The slight glass swelling (about 0.5%) was linked to an appreciable improvement in its mechanical properties, i.e. a 30% hardness reduction and increased resistance to cracking [Peuget et al. 2006]. By molecular dynamics simulation of the accumulation of radiation induced collision cascades in a sodium borosilicate glass Bureau et al. (2008) showed that the mechanism underlying swelling is related to atom ejection and thermal quenching.

The effect of γ-radiation on wastes was studied up to a dose of 8 MGy showing no effect on glass structure (McGann et al. 2012). In contrast, radiation may influence oxidation state of redox sensitive elements in the glass: For Gd containing glasses EPR spectroscopy has shown that under β-irradiation the relative content of Gd occupying a network former position increased compared to Gd occupying a network modifiers site. The evolution of the EPR lines appears to result from the reduction processes of $Gd^{3+}$ into $Gd^{2+}$ [Maluchkova et al. 2006].

**The glass dissolution process in groundwater**

The corrosion of radioactive and inactive borosilicate nuclear waste glasses has been studied for more than 30 years. Studies aimed at determining leach rates from glasses have resulted in computerized models for predictive purposes, at identification of surface reaction products and of dissolved glass constituents. Glass alteration is a surface reaction processes coupled with diffusion processes of reactants (water molecules) and dissolved glass constituents (dissolved silicic acid…). In a first reaction step water molecules diffuse into a subsurface of the glass thus creating à hydrated surface layer. Alkali ions are mobile in this layer and by ion exchange with protons from the aqueous solution the surface becomes depleted from alkali ions. The main glass constituents like silica remain in the hydrated glass and silanol groups formed by alkali ion exchange condense to form a stable $SiO_2$ rich surface. Further progress of the reaction leads to a transformation of the hydration reaction into a gel like silica rich reaction product. The gel can be considered as a very porous solid solution of hydrated sparingly soluble phases (oxides hydroxides, silicates…) of glass constituents. Solution constituents take part in gel formation. The composition of the gel can be calculated by solid solution models implemented in geochemical codes (Munier et al. 2004). The gel layer provides also a very high density of sorption sites for radionuclides all along the interior of this porous medium. Often the gel does not form a single solid solution, but gel formation is accompanied by the formation of crystalline phases which form at the outer glass surface (like clay minerals) but sparingly crystalline phases can form also within the gel. Typical crystalline phases are powellite, barite, calcite, anhydrite and above all of clay-like Mg(Ca,Fe)-silicates [Bosbach et al. 2009].

The dissolution or corrosion rate of the glass matrix in ground water provides an upper limit for the release rate of uranium from the glass matrix, but the incorporation of uranium in secondary phases may provide much lower rates of release. It is the rate by which the front of gel formation moves into the glass. The dissolution rate of the glasses in pure water decreases with time. Very low values of the dissolution rate are achieved under conditions of solution saturation. The term "saturation" refers here to the achievement of a maximum value of the activity of ortho-silicic acid in the aqueous solution, typically in the milli-molar range. It does not represent a true thermodynamic equilibrium and glass dissolution continues, all though at a very low rate. Achievement of saturation is expected for most geological environments since groundwater volumes are small. This decrease in dissolution due to saturation results in longer lifetimes for the vitrified waste, and so contributes to the greater long-term safety of geological disposal.

The understanding of the long-term dissolution rate and its precise value are still the subject of debate. Two explanations for the decrease in dissolution rate have been considered. The decrease can arise as the solution concentration of $H_4SiO_4$ approaches a solubility limit thereby decreasing the driving force for the dissolution (affinity concept). The second explanation is that the secondary layers that develop on the surface of the glass act as transport barriers ( the protective layer concept). Both interpretations refer to concepts that were proposed as early as the 80's. The European commission project GLAMOR helped to resolve remaining conceptual uncertainties [Van Iseghem et al. 2009]. There was agreement that both affinity and protective layer concepts must be taken into account to explain long term glass dissolution.  There is a consensus that a slow residual dissolution rate exists and that potential solid/water equilibrium will never lead to situations where glass dissolution rates become zero.  This residual rate significantly controls the long-term release of very soluble elements like iodine or selenium from the glass, and it is an important parameter in the safety assessment for final disposal. Various processes have been invoked to account for this process: water diffusion, ion-exchange reactions, network hydrolysis, and precipitation of secondary phases.

**Solubility controls in actinide release from glass to groundwater**

Glass dissolution rate related data and the corresponding mechanistic understanding are limited to predicting the stability behavior of the glass matrix which does not describe the often much lower rates of release of sparingly soluble radioactive glass constituent oxides. Solution concentrations of sparingly soluble glass constituents are principally controlled by thermodynamic instead of kinetic constraints. These constraints determine how much actinides will remain bound in the surface layer of solid reaction products on the glass (gel and/or crystalline phases). But the thermodynamic constraints are often difficult to assess, since very often solid solutions are formed and solubilities are controlled by poorly crystalline phases for which thermodynamic data are rare. In case of solid solution formation, composition of this solid may be variable with the progress of the glass dissolution reaction as has been show for an Am containing powellite solid solution phases (Grambow et al. 1996). Furthermore, weak solubility and poor crystallinity of solubility controlling phases have the drawback that these phases are difficult to characterize by solid state or surface analytical techniques such as X-ray diffraction (XRD), transmission electron microscopy (TEM) etc.. Additionally, the solubility controlled release of uranium from the glass is highly dependent on many variables, processes and boundary conditions which depend also on the progress of glass dissolution. The reason is that during glass dissolution, the composition of the aqueous solution and its pH vary with time. For example, a typical groundwater collected at a potential repository site may have a pH of around 7. However, due to glass dissolution, this groundwater pH may be shifted towards a value of 9. Thermodynamic solubilities may be quite different at pH 7 than at pH 9. Hence it is important to study solubility constraints of sparingly soluble radionuclides in the context of glass dissolution studies, for example by using experimental solubility's for uranium released from pulverized uranium doped glasses. Such studies have been performed in a wide range of well-controlled conditions, considering important variables such as pH and Eh of the groundwater or leach solutions, and their comparisons at the given aqueous composition. A detailed assessment of the impact of solubility constraints on radionuclide release from nuclear waste glasses has recently been provided by Rai et al. (2011). Experimentally measured solution concentrations were compared with theoretical solubilities predicted for known solid phases from known thermodynamic data. These comparisons can then be used to indirectly infer the specific solubility-controlling solids in actinide containing glass/water systems when a direct determination by solid state analytical techniques is impossible. However, this requires that all thermodynamic data for all relevant solution species and corresponding activity coefficient correction are known for the temperature of interest and that also concentrations of potential complexants in the aqueous solution are known. These constraints are not always given: In the case of controls for release of Th from borosilicate glass, it appeared clear that a $ThO_2$(am) phase controls release for any pH larger than 1, but detailed thermodynamic solubilitiy calculations could not be made since no thermodynamic data were available for the observed [Rai et al. 2005] dominant Th silicate solution complex. Rai et al. (2011) showed that the solubility-controlling solids of neptunium may include $NpO_2$(am), $NpO_2$(cr), and $(An_x, Np_{1-x})O_2$(s) with "An" = actinide, whereas the solubility-controlling solids of plutonium may include $PuO_2$(am), $PuO_2$(cr), and $(An_x, Pu_{1-x})O_2$(s). Above pH 4, all observed solution concentrations seems to be solubility controlled. However, under reducing conditions, Pu(III) will be the dominant oxidation state. Pu(III) is known to behave in a manner analogous to other trivalent actinides and rare earths. For the release of trivalent actinides, glass dissolution kinetics controls the release for pH values lower than 5.5 whereas solubility constraints become dominant at higher pH with possible solubility controlling phases being $(REE,An)(OH)_3$(s) solid solutions. Solid solution formation is very important in this system, since the nuclear waste glass contains in most cases much more trivalent rare earth elements than trivalent actinides. Formation of $(REE, An)OHCO_3$(s) phases seems also be possible. Thermodynamic calculations are also strongly dependent on the presence of phosphate in the glass.

**Measures of progress of glass corrosion**

The "switch" between kinetic and solubility control of actinide release depends on the progress of the dissolution reaction. The solution concentrations of dissolved glass components are a good indicator of the progress of the glass / water reaction if the following conditions are met: (1) it is a static test (2) the dissolved glass components remain in solution and do not participate in secondary precipitation or sorption reactions, (3) the surface area of the glass sample is known. The condition (2) is valid for soluble glass constituents such as boron, but it is often met for many other elements. It has proven to be useful to normalize the solution concentrations to the concentration of the element in the glass and on the ratio of the glass surface area (S) exposed to a given volume (V)of solution (groundwater or other leachant). Thus we obtain the so-called normalized elemental mass loss NL of the element i from the glass. $NL_i$ is calculated from the measured concentration $C_i$ of i in solution based on the weight fraction $f_i$ of i in glass

$$NL_i = \frac{C_i}{f_i \cdot (S/V)}$$

were S is the surface area of the glass and V is the volume of the aqueous solution in a static leach test. Typical units for $NL_i$ are in [g/m$^2$] . Dividing the elemental mass loss value NL by the specific weight of the glass used in the test gives the equivalent average elemental depletion depth d of glass. This depletion depth corresponds to the thickness of a glass surface layer, which should be dissolved to account for the measured solution concentrations of the element i. Calculated from the solution concentrations of boron, corresponding equivalent thicknesses often correspond to a first approximation of the measured thicknesses of corrosion layers on the glass surface.

In certain cases it is more useful to normalize the progress of the glass water reaction not to the glass surface area but to the solution volume. This is particularly the case if the effect of solution saturation and solubility of secondary phase formation on glass dissolution has to be studied. A reaction progress indicator $\xi$ can easily be obtained by multiplying the $NL_i$ value of a soluble element like boron by the S/V ratio: $\xi=NL_B*S/V$. It is the concentration of boron, normalized to the weight fraction $f_B$ of boron in the glass. Similar normalized concentrations $NC_i$ can be obtained for all other elements measured in aqueous solution ($NC_i=NL_i*S/V$). In case of solubility control, NC values may remain constant, they increase slower than the NC value of boron or they may even decrease.

**Release behavior of actinides from glass**

The analyses of aqueous solutions resulting from glass leaching experiments provide quantitative data on the retention of major as well as trace elements. Often inactive simulations of the actual nuclear waste glasses are used, indicating that information on retention is to some extent indirect as for example, the behavior of trivalent actinides (Am, Cm) and Pu is often deduced by analogy with the trivalent lanthanides and Th/U, respectively. Retention factors well above 99% were observed for example for all lanthanides for static leaching experiments of a Mg rich borosilicate glass (Magnox waste MW glass) at high S/V ratios and of 12 years of duration [Curti et al. 2006]. Reasons for the strong retention of trivalent elements are the formation of rather insoluble oxyhydroxides but as well of strong sorption. Curti et al. (2012) have shown by spectroscopic techniques (µ-XRF, XANES) that Ce is bound as Ce(III) on Mg rich corrosion products. Sorption of trivalent lanthanides on glass corrosion products was found much stronger than on known smectite clays [Luckscheiter et al. 2006]. EXAFS studies have shown that the lanthanide aquo ions $Ln(H_2O)_{8-10}^{3+}$ are incorporated by intercallation between the interlayers of the smectite clays formed also by glass corrosion [Monoz-Paéz et al. 1995]. Only at high temperature of dehydration or long drying times conversion into the hydroxide structure $Ln(OH)_3$ may occur. On the other hand, Time resolved laser fluorescence spectroscopic (TRLFS) measurements on Eu doped hectorite suggest the structural uptake of lanthanides (and trivalent actinides) into the octahedral layer of clay minerals [Bosbach et al. 2009]

The main pentavalent radionuclides under oxidizing conditions are neptunium (V) and plutonium (V). In the presence of oxygen atoms (glass, water, gel) Np and Pu are present as a singly positive charge neptunyl or plutonyl ion of rather high mobility. In the leaching of HLW glass under oxidizing conditions therefore only a rather low retention of Np (V) in the surface layer observed [Vernaz 1992].

Pu is strongly retained in the glass alteration products during glass leaching. Lanza et al (1982) observed 3 orders of magnitude lower Pu release rates compared to Cs release. Pu release and its temperature dependency appears to be affected by the solubility of $PuO_2 \cdot xH_2O$ remaining in the leached surface layers [Banba et al. 1989]. In Lanthanide-Borosilicate waste glass with 7 or 15wt% $PuO_2$ loading, release of Pu was much slower than that of boron, which is used as indicator for the dissolution of the glass matrix [Bibler et al. 1996]. Despite the fact that the 15wt% glass shows phase separation of $PuO_2$, no increased leaching was observed, indicating again, that the release rate of Pu is not controlled by the rate of glass dissolution but by secondary phases.

A plot of the experimentally measured values of $NC_i$ vs. $\xi$ allows assessing the onset of sorption and solubility constraints on the release of elements of interest from the glass. For soluble non-sorbed elements $NC_i = \xi$, whereas $NC_i < \xi$ in case of sorption or precipitation. Figure 1 gives an example for a typical nuclear waste glass (SON68), leached at 190°C in highly saline aqueous solutions (Q-brine) with NaCl=11,03 g, MgCl$_2$ x 6H$_2$O = 436,12 g, KCl=36,01 g, MgSO$_4$ x 7H$_2$O=21,89 g and H$_2$O=257 ml. Experimental results of various static tests for experimental durations up to 4 years and for ratios of surface to volume ratios S/V between 10 and 21000 m$^{-1}$. The experimental conditions are detailed in Grambow et al.(1988, 1990). Homogeneous curves were obtained for most elements. This indicates that solution concentrations are a function of reaction progress. Normalized Li concentrations are in most cases higher than those of boron, indicating selective release (alkali ion exchange). For large ranges of reaction progress elements like uranium, calcium and strontium follows the boron curve, indicating congruent release proportional to the inventory of these elements in the glass, and absence of solubility constraints. However at reaction progress higher than about 5000 g/m$^3$ uranium and neodymium concentrations do not rise anymore, indicating some kind of solubility control. At about the same reaction progress, also Mo reaches limiting solution concentration values. For comparison, silicic acid and Al concentrations have reached saturation values at much lower concentrations. In the case of Nd, formation of alkali-rare earth molybdates has been reported as potential solubility controlling phase for the given experimental conditions [Grambow et al. 1991].

Figure 1: Evolution of normalized solution concentrations in glass dissolution experiments (saline solutions at 190°C, see text) as a function of the reaction progress

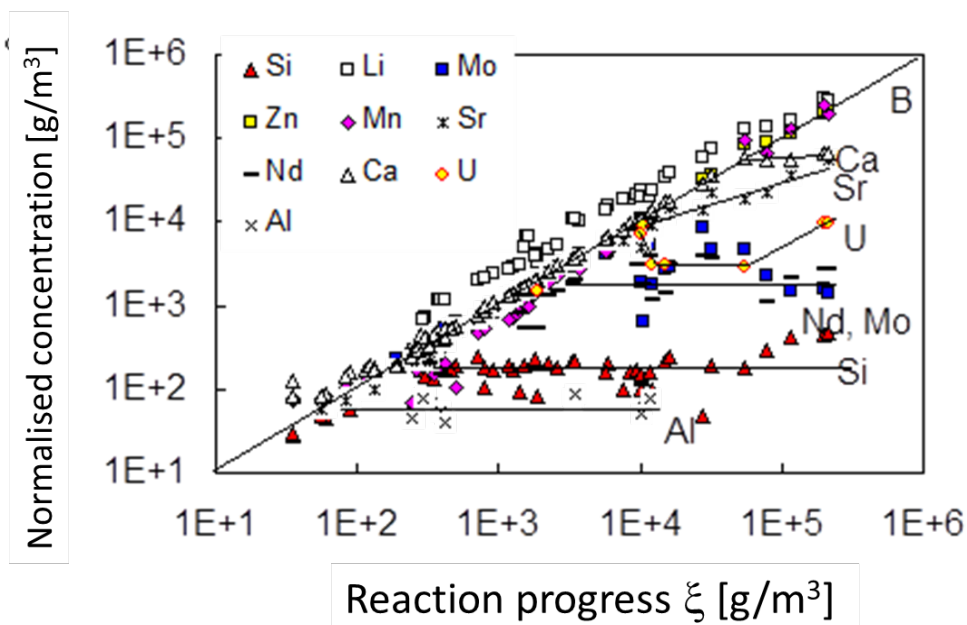

It is interesting to study the temperature dependency of redox state distribution of Pu in these Mg rich acid saline solutions: by leaching Pu, Np; Tc and Am doped glasses between 110 and 190°C Pu(V,VI), were found most abundant in leachates from experiments at 110°C. Leachates from experiments at 190°C contained Pu(III) and Pu(IV) and no Pu(VI). Distribution of Pu between surface layers and aqueous solutions correspond to these redox states: at 110°C Pu is relatively mobile leading to a situation where the half of the Pu inventory of the dissolved glass is found in aqueous solution. At higher temperature about 99% of the Pu is retained in the glass corrosion products. The trivalent americium is always retained to at least 83% in the leached layer, with higher rentention in the less acid leachates obtained at lower temperature [Grambow et al. 1995, Grambow et al. 1996].

Actinides are often found incorporated as minor component in major glass corrosion product. For the experiment with the leach data described in figure 1, figure 2 shows the sequence of crystalline phases observed with reaction progress. The figure was derived from original data reported by Grambow et al. 1990, Abdelouas (1996) and Abdelouas et al. (1997). Zircon contained 5 wt% of Nd and 7 wt% of Th. Powellite (nominally $CaMoO_4$, but large quantities of Ca were replaced by Na and a trivalent rare earth element) contained 23wt% of Nd, 6% Pr and 4% La and similar incorporation of trivalent actinides has been predicted [Grambow et al. 1996).

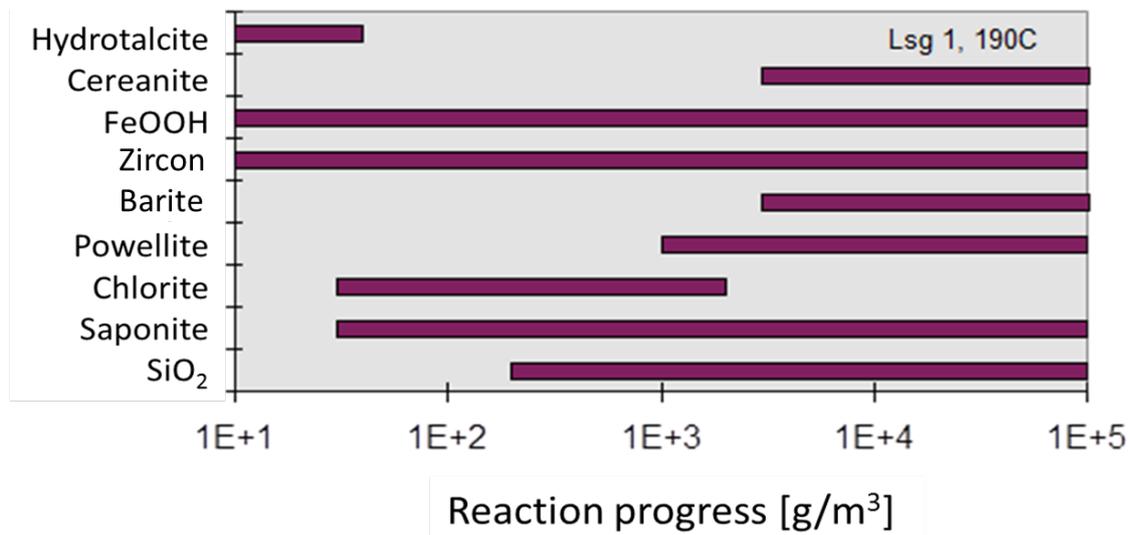

Figure 2: Evolution of reaction products of SON68 glass dissolution in Mg rich brines as a function of reaction progress

Leaching of simulated uranium doped nuclear waste glass in deionized water at 90°C under static conditions and a low S/V ratio of 10 $m^{-1}$ led to measured solutions concentrations of uranium of $2.5·10^{-6}$ mol/L, closely similar to the predicted solubility of $UO_2(OH)_2$ [Maeda et al. 2001]. Normalized concentrations were reported as 100 $g·m^{-3}$, about 10 times lower than those of Na. Uranium was accumulated in the 60μm thick alteration layer on the glass surface, but there was no or very little incorporation of uranium in fibrous clay-like phases, consistent with formation of uranyl oxyhydrate phases. Under such conditions, uranium release from the glass is expected to be controlled by solubility of secondary phases rather than by the kinetic stability of the glass. Normalized U concentrations at saturation are about 50 times lower than those presented in figure 1. The reason for the high solubility in saline solutions is the acid pH value of these Mg rich solutions.

**Effect of water flow rate on actinide behavior**

In leaching experiments in fast flowing water (Single Pass Flow Through or SPFT test, water residence time was <1 day) at 88°C with the plutonium doped DRG-P glass [Wellman et al. 2004] normalized uranium release rates between pH 9 and 12 were similar than those of Al, B, and Cs, indicating release rate control by congruent glass matrix dissolution. Indeed, since water flow rates were very high and glass dissolution is slow, glass dissolution controlled solution concentrations of uranium (and of other glass constituents) become lower than the solubility limiting values. For example, at pH

9 observed uranium concentrations were in the range of $3 \cdot 10^{-7}$ mol/L, lower than the solubility of potential U(VI) solid phases.

Jollivet et al. (2005) observed that the coordination number of Ce(IV) increased in the gel layer of the glass compared to its environment in the glass. As a general rule, the gels obtained with a high leachate renewal rate (high water flow rate) exhibit higher total coordination numbers than ones obtained at low flow rate. From Ce-O distance in the gel it was deduced from glasses with low Ce contents that Ce exists both in silicate and in oxide environments in the gel. At high water residence times (high silica concentrations in the aqueous solution) silicate sites are favored, at low water residence times, the oxide sites. Indeed, gels formed at high water renewal rates had lower silica contents. At high Ce contents of the glass and independent of water renewal rates, the coordination environment of Ce in the gel shows strong Ce-Ce contributions, indicating formation of Ce rich phases. Nd exists in phosphate free glass only in a silicate environment, while in the gel both silicate and oxy-carbonate environments were observed, the ratio between both environments decreasing with increasing water renewal rate.

**Colloid formation and actinide release**

If actinide release from glasses is controlled by thermodynamic solubility of secondary phases, release rates are very low for tri- and tetra-valent actinides since the corresponding solubility's of secondary phases are low. On the other hand radionuclide release from glass can become much higher than solubility controlled values if colloids are formed. The effect is more pronounced in typical groundwaters (smectite colloids) than in saline brines [Geckeis et al. 1998]. For the corrosion of glass COGEMA R7T7 in distilled water it was observed that only about 3 ‰ of the Am content on the corroded glass is mobile, while 99.7% remained undissolved in the surface layer of the glass. The mobile fraction consisted largely of colloids [Vernaz et al. 1992]. One very important variable is the carbonate content of the water used for the dissolution experiments, because by forming Am carbonato complexes the solubility of Am can be increased substantially. Thus, the leaching of Am in degassed distilled water is lower than in air-saturated water containing $CO_2$. Also in Pu containing glasses approximately 80% of the total measurable Pu was in the form of a colloidal particles in a size range between 1.8 and 200 nm, hence Pu is migrating principally as or in association with colloids. In natural environments, organic matter is present (humic acids) in groundwater and may strongly alter colloid formation [Wang et al. 1992].

A detailed study of the colloid formation mechanism during sodalite glass dissolution has been presented by Morss et al. (2001). TEM analyses of colloids from glass leaching showed substantial amounts of ~10 nm aluminosilicate colloids or clusters of these colloids and smaller amounts of colloidal crystalline $(U,Pu)O_2$. The $PuO_2$ particles that were found associated with aluminosilicate aggregates. X-ray absorption spectroscopy showed that plutonium in the colloids is tetravalent and has the coordination geometry of $PuO_2$ indicating that plutonium contained in the glass is released during corrosion as $(U,Pu)O_2$ colloids.

**Redox behavior of actinides and rare earth elements at glass/water interfaces**

At glass/water interfaces uranium may exist either in tetravalent or hexavalent state. Factors which influence the oxidation state are the environmental boundary conditions, the redox state of the glass melt in the production process as well the presence of engineered barrier materials during disposal, such as corroding containers made of iron based materials. The chemical environment of the retained radionuclides in the surface layer of the glass may be different from that in the original glass phase. This is particularly evident in the exchange of the coordination number of the ions, which can be studied with the help of EXFAS or XANES. For example $Mn^{2+}$ in the dry glass is often present in four-fold coordination bound to the glass (network former), while the coordination number is increased in the hydrated glass surface to six (network modifier) [Kohn et al 1990]. $Sr^{2+}$ also increases the coordination number in the glass from 6 to 7 at the surface.

Uranium readily oxidizes to the hexavalent state in the absence of extremely reducing conditions. The manner in which the glasses are produced exhibits often fairly oxidizing conditions such that the dominant oxidation state of U in glasses is expected to be U(VI). But redox states during glass melting can be adapted using tailored Fe(II)/Fe(III) ratios in the inactive glass frit, or alternatively formic or oxalic acid and Ar bubbling. Repository near field conditions can be extremely reducing caused by hydrogen production due to container corrosion. Most known U(VI) solids are soluble with solution concentrations in the range of $10^{-6}$ to $10^{-4}$ Mol/L compared to U(IV) solids with solubilities in the range of $10^{-8}$ mol/L.

Curti et al. (2012) observed Ce(III) in the gel layer of the glass even though Ce was tetravalent in the glass phase. In contrast, Jollivet et al. (2005) and Yang et al. (2006) observed Ce(III/IV) in the glass, but formation of Ce(IV) in the corroded glass surface layer. Obviously redox conditions in glass making and during leaching were different in these two experiments. Curti et al. (2012) invoked the stainless steel reaction vessel corrosion as source for reduction of Ce(III).

**Actinide release under oxidizing conditions**

Uranium concentrations in oxidizing glass leachates will most likely be governed by the presence of oxidized uranium (U(VI)) and its hydrolysis and complexation behavior in aqueous solution. Leaching of the glass SON68 with 80% U(VI) and 20%U(IV) under oxidizing conditions in deionized water has led to complete oxidation of the U(IV) content of the corroded glass. In the glass both uranate and uranyl environments were present while in the gel layer only the uranyl environment remained [Ollier et al. 2003]. At fast water renewal rates (low normalized solution concentrations) in the gel uranyl hydroxide environments were detected, while at low renewal rates uranophane was observed, combining XPS and time resolved photoluminescence spectroscopy [Ollier 2002]. Probably, at high water renewal rates, calcium and silica concentrations remained too low to allow for uranium silicate formation.

Hexavalent uranium forms in the presence of oxygen (glass, gel, water) uranyl ions, which are coordinated in the glass network or in the gel layer by two axial oxo groups and by four or more equatorial oxygen atoms. While the uranyl ion in the glass is planar coordination with four equatorial oxygen atoms, the coordination number is increased in the aqueous surface layer of corroded glasses. Simultaneously an increasing tendency with time for clustering of uranyl units in the surface layer of the glass was detected, which can be interpreted as a precursor to the formation of uranyl oxyhydrates. For the corrosion of uranium-rich HAW glasses over a sufficiently long reaction times at a sufficiently high ratio of glass surface area to solution volume one can detect different uranyl silicates such as weeksite ($K_2(UO_2)_2Si_6O_{15}\bullet4(H_2O)$), formed as secondary phases [Barrett et al. 1986]. Other reaction products are the uranyl silicates uranophane ($Ca(UO_2)_2SiO_3(OH)_2\bullet5(H_2O)$) and haiweeite($Ca[(UO_2)2Si_5O_{12}(OH)_2]\bullet3(H_2O)$)  [Ebert et al. 1991]. Uranophane was also observed in the the gel layer of U containing borosilicate glass by Ollier (2002). Additionally bauranoite ($BaU_2O_7\cdot xH_2O$) was observed.

In aqueous solution hexavalent uranium forms very stable highly mobile negatively charged carbonato complexes (eg, $UO_2(CO_3)_3^{4-}$) [Wronkiewicz et al. 1993]. These lead to high solubility of any type of U(VI) containing secondary phase. Some studies were performed to take typical glass leachate compositions and to add uranium until solubility limits are reached [Ribet et al. 2004]. Such glass leachates spiked with high concentrations of U(VI) indicate that schoepite ($UO_3\cdot2H_2O$) may be the solubility-controlling phase. Depending on the leachate composition, other potential solubility controlling phases were U(VI)-phosphate, becquerelite ($Ca(UO_2)_6O_4(OH)_6\bullet8(H_2O)$), or uranates. In contrast under more realistic conditions in presence of a glass phase, the similarity of observed U(VI) concentrations to the calculated concentrations in equilibrium with schoepite ($UO_3\cdot2H_2O$) [Luckscheiter et al. 2006] does however suggests that under oxidizing conditions schoepite may be the solubility-controlling solid.

**Actinide release from glasses under reducing conditions**

The presence of large quantities of iron in a nuclear waste repository leads to reducing conditions to prevail where in U(IV) is dominant and potentially existing U(VI) is expected to convert to the most insoluble U(IV) oxidation state. Importantly other radionuclides that may be present under such conditions in the tetravalent state, the actinide plutonium, neptunium,  thorium, and the fission product technetium. The tetravalent actinides behave rather similar to tetravalent uranium, hence, we can use data for these elements to deduce on the behavior of uranium(IV)

When the corrosion of COGEMA glass R7T7 was tested in distilled water [Vernaz 1991], only about 2% of the Pu content of the corroded glass remained  mobile, while 98% remained insoluble in the surface layer of the glass. The mobile fraction consisted largely of colloids. In the very fast flowing water of a boiling Soxhlet-test(no saturation of dissolved silica, very dilute water, no accumulation of corrosion products) only 2 ‰ of the Pu(IV) content of the corroded glass was mobilized. In natural glasses containing uranium, thorium and zirconium were studied. Uranium, besides its tetravalent state exists often in oxidation state VI and seldom V. The thorium and zirconium oxidation under natural conditions are stable only in the tetravalent form.

In corrosion experiments [Claparols 1992] of volcanic glasses exposed to different media (fresh water, sea water) and under different conditions (low temperature and hydrothermal), tetravalent elements like Th and Zr remained entirely immobile  during the corrosion process and  the ratio between these two elements at the glass surface were similar before and after the corrosion process. By geochemical mass balance of  weathering of icelandic  hyaloclastites (natural basaltic glasses) under natural geological conditions (100000 of years) an absolute release of rare earths and thorium was observed which was higher in the early stages of glass corrosion (small ratio of glass surface area to solution volume) which became lower, the more the reaction had progressed.

Studies [Bosbach et al. 2009] of the simulated German nuclear waste glass GP WAK1 (composition comparable to the COGEMA R7T7 glass) in powdered form at S/V= 5000 m$^{-1}$ in the two synthetic clay water solutions both with and without added NaHCO$_3$ and with added cast iron chip to simulate reducing conditions gave normalized uranium concentrations NL$_U$*S/V of 5 g/m$^3$ (actual concentrations are between 1 and 3 10$^{-7}$ mol/L for a boron based reaction progress of 10000 g/m$^3$ (data derived from the figure 2 of the authors), about 1000 times less than given for the data in figure 1. Even though fully reducing conditions were not attained (the Eh remained at 140 mV) the presence of the cast iron chips seems to have led to low uranium concentrations despite high reaction progress of glass dissolution. Leaching under these conditions did not lead to reduction of U(VI) to U(IV) in the glass surface since the oxidation state of the original glass (VI) remained unchanged. Instead the coordination number of equatorial ligand oxygen atoms of the uranyl ion increased from 4 in the glass to 5 in the leached layer (see also table 1).

Studies with powdered iron with high surface area [Xia et al. 2011] and U(VI) solutions showed that U(VI) is reduced to the most insoluble U(IV) oxidation state and precipitated as UO$_2$(am). Solubility studies [Rai et al. 2003] with UO$_2$(am) where iron powder was used to maintain the tetravalent state showed that UO$_2$(am) relatively readily converted to UO$_2$(cr) at moderately high temperatures (90 °C in this case). The conversion kinetics of UO$_2$(am) to UO$_2$(cr) at expected repository temperatures (50 to 60 °C) are not known. However, judging from the temperature effects at 90 °C, it can be stated that the rates of these conversions should be relatively faster at repository temperatures than at room temperature. Therefore, even though specific data on solubility-controlling solids in glasses are not available, assuming that reducing conditions will be maintained, UO$_2$(am) in a most conservative scenario can be used as a solubility-limiting phase. However, it should be mentioned that total U concentrations compared to other tetravalent actinides will be high (high mole fraction) and thus the U(IV) concentrations in equilibrium with the solid solution (An, U)O$_2$(s) will not be significantly different than the pure phase UO$_2$(s).

**Actinide release from glass under field conditions**

The controls of actinide release from the glass become more difficult to assess under field conditions. Here, solution concentrations of uranium may remain low due to sorption on adjacent minerals. Stability of uranium colored historical glass buried at 50 cm depth for 150 years in soils and exposed to leaching by soil solutions under natural conditions for 150 years was studied [Prochaska 2007]. Uranium in the glass is bound much more firmly than alkalis, which were found to readily leach. Even if the glass lost 14% of its mass by leaching, the uranium content in the leached surface layer increased. There could be an apparent increase of U concentrations in a leached layer, caused by normalizing all elemental contents in the gel layer to 100%, but there may also be an absolute enrichment of U in the leached layer, caused by uranium sources different from that of the glass. Only by detailed isotopic studies can one distinguish between these cases.

By geochemical mass balance of weathering of Iceland-hyaloclastites (natural basalt glasses) under natural geological conditions (100,000 of years) an absolute release of rare earths and thorium in the early stages of glass corrosion (bulk glasses, small ratio of glass surface area to solution volume) was observed [Daux et al. 1994], which again slowed down the more the reaction progressed. Here, the rare earths were found attached in the secondary, poorly crystallized phases (Palagonites) rather than in the better-crystallized phases.

**Conclusion**

Borosilicate glass can be an important confinement matrix for immobilization of actinides from nuclear wastes in geological disposal environments. However only in fast flowing ground water or under conditions of very high actinide solubility (acidic Mg rich brines) will the glass matrix alteration limit the release of the actinides to the groundwater. In typical geological disposal environments, ground water flow and U solubilities are low. Under these conditions, secondary phases control actinide release. Under oxic conditions, the solubility-limiting solids are expected to be the U(VI) solids, amorphous or crystalline $PuO_2$, $Am(OH)_3$ etc.. Under reducing repository conditions the secondary solid phase that will limit U concentrations, depending on repository temperatures, is most likely $UO_2(am)$, $UO_2(cr)$ or if other tetravalent actinides are also present then one would expect the solubility-limiting phase to be the ideal solid solution $(An,U)O_2(s)$. Pu may be bound to trivalent hydroxide phases. If secondary phases control actinide concentrations at the glass/solution interface, the safe isolation of actinides from the biosphere depend only to a very small degree on glass properties such as glass dissolution rates. More important is the interplay of thermodynamic constraints of solubility controlling actinide containing solid phases with geochemical parameters such as carbonate concentrations, solution pH or hydrodynamic parameters such as ground water flow rates.

**References**

A. Abdelouas, Le mécanisme de corrosion en phase aqueuse du verre nucléaire R7T7. Approche expérimental. Essai de modélisation thermodynamique et cinétique. PhD thesis, Université Louis Pasteur, Strasbourg, France (1996)

A. Abdelouas, J.L. Crovisier, W. Lutze, B. Grambow, J.C.Dran, R. Müller, « Surface Layers on a borosilicate glass corroded in MgCl2 solution », J. Nucl. Mater. 240 100-111 (1997)

M. Antonini, A. Merlini und F.R. Thornley, in "EXAFS and Near Edge Structure III" Springer Proceedings in Physics, Vol 2. Eds. K.O. Hodgson et al. (1984) 349

T. BANBA, K. NUKAGA and T. SAGAWA, "Temperature Effect on Plutonium Leach Rate of Nuclear Waste Glass," Journal of Nuclear Science and Technology, 26 (1989) 705-711

N.T. Barrett, G.M. Antonini, G.N. Greaves, F.R. Thornley und A. Manara, Journal de Physique, Colloque C8, Suppl. No. 12, Tome 47 (1986) C8-879-882.

N. E. Bibler, W.G. Ramsey, T.F. Meaker and J.M. Pareizs, "Durabilities and Microstructures of Radioactive Glasses for Immobilisation of Excess Actinides at the Savannah River Site," Mat. Res. Soc. Symp. Proc. 412 (1996) 65-70

D. Bosbach, B. Luckscheiter, B. Brendebach, M.A. Denecke, N. Finck, "High level nuclear waste glass corrosion in synthetic clay pore solution and retention of actinides in secondary phases," Journal of Nuclear Materials 385 (2009) 456–460.

O. Bouty, J. M. Delaye, S. Peuget, "Europium structural effect on a borosilicate glass of nuclear interest, "Procedia Chemistry 7 ( 2012 ) 540 – 547

G. Bureau, J.-M. Delaye, S. Peuget, G. Calas, "Molecular dynamics study of structural changes versus deposited energy dose in a sodium borosilicate glass," Nuclear Instruments and Methods in Physics Research B 266 (2008) 2707–2710

J.-N. Cachia, X. Deschanels, C. Den Auwer, O. Pinet, J. Phalippou, C. Hennig, A. Scheinost, "Enhancing cerium and plutonium solubility by reduction in borosilicate glass," Journal of Nuclear Materials, 352 (2006) 182–189

Ch.Claparols, Comportement des éléments en traces et variations des compositions isotopiques du Nd et du Sr au cours de l'altération de dépôts volcaniques: études de cas et application aux séries mésozoïques de black shales de l'Océan Austral. PhD thesis, Université Toulouse (1992)

E. Curti, J.L. Crovisier, G. Morvan, A.M. Karpoff, Long-term corrosion of two nuclear waste reference glasses (MW and SON68): A kinetic and mineral alteration study, Applied Geochemistry, 21 (2006) 1152–1168

E. Curti, D. Grolimund, C. N., Borca. A micro-XAS/XRF and thermodynamic study of CeIII/IV speciation after long-term aqueous alteration of simulated nuclear waste glass: Relevance for predicting Pu behavior? Applied Geochemistry 27 (2012) 56–63

V. Daux, J.L. Crovisier, C. Hemond und J.C.Petit, Geochim. Cosmochim. Acta 58, (1994) 4941-4954

X. Deschanels, S. Peuget, J.N. Cachia, T. Charpentier, Plutonium solubility and self-irradiation effects in borosilicate glass, Progress in Nuclear Energy 49 (2007) 623-634

W.L. Ebert, J.K. Bates und W.L. Bourcier, Waste Managment 11, (1991) 205-221

F.A. Farges, G.E. Brown und C.W. Ponander, EXAFS study of the structural environments of trace levels of Zr4+, Mo6+ and U6+/U5+/U4+ in silicate glass/melts systems, in "X-Ray Absorption Fine Structure" Hasnain ed., Ellis Horwood, New York (1991) 309-311

J.A. Fortner, E.C. Buck, A.J.G. Ellison, J. K. Bates, EELS analysis of redox in glasses for plutonium immobilization, Ultramicroscopy 67 (1997) 77-81

H. Geckeis, B. Grambow, A. Loida, B. Luckscheiter, E. Smailos, J. Quinones, « Formation and stability of coloids under simulated near field conditions », Radiochimica Acta, 82, 123-128 (1998)

B. Grambow, W. Lutze, R.C. Ewing, L.O. Werme, "Performance Assessment of Glass as a Long-Term Barrier to the Release of Radionuclides into the Environment", Mater. Res. Soc. Symp. Proc. 112, 531-542 (1988).

B. Grambow, R. Müller, "Chemistry of Glass Corrosion in High Saline Brines", Mat. Res. Soc. Symp. Proc. 176, 229-240 (1990)

Grambow B., R. Müller, A. Rother and W. Lutze, "Release of Rare Earth Elements and Uranium from Glass in Low pH High Saline Brines", Radiochimica Acta, 52/53, 501-506 (1991)

Grambow, B., A. Loida, L. Kahl, W. Lutze, "Behavior of Np,Pu,Am,Tc upon Glass Corrosion in a Concentrated $Mg(Ca)Cl_2$-Solution"; Mat. Res Soc. Symp. Proc. 353, 39-46 (1995)

B. Grambow, W. Lutze, L. Kahl, H.Geckeis, E.Bohnert, A.Loida, P. Dressler, R.Pejsa, "Retention of Pu, Am, Np and Tc in the Corrosion of COGEMA Glass R7T7 in Salt Solutions" Final Report of EU-Project, Wissenschaftliche Berichte, FZKA 5767, Forschungszentrum Karlsruhe (1996) Commission of the European Union : EU 17114 EN

G.N. Greaves, J. non-cryst. Solids,, "EXAFS for Studying Corrosion of Glass Surfaces," 120,108-116 (1990)

N.J. Hess, W.J. Weber, S.D. Conradson, J. Nucl. Mater. 254 (1998) 175.

P. Jollivet, C. Den Auwer, E. Simoni, Evolution of the uranium local environment during alteration of SON68 glass, Journal of Nuclear Materials 301 (2002) 142–152

P. Jollivet, C. Lopez, C. Den Auwer, E. Simoni, Evolution of the local environment of cerium and neodymium during simplified SON68 glass alteration, Journal of Nuclear Materials 346 (2005) 253–265

A.Kidari, M.Magnin, R. Caraballo, M.Tribet, F.Doreau, S. Peuget, J-L. Dussossoy, I. Bardez-Giboire, C.Jégou, Solubility and partitioning of minor-actinides and lanthanides in alumino-borosilicate nuclear glass, Procedia Chemistry 7 ( 2012 ) 554 – 558

G. S. Knapp, B. W. Veal, A. P. Paulikas, A. W. Mitchell, D. J. Lam, and T. E. Klippert, EXAFS STUDIES OF SODIIM SILICATE GLASSES CONTAINING DISSOLVED ACTINIDES, CONF-840764 DE84 015835 (1984)

S.C. Kohn, J.M. Charnock, C.M.B. Henderson und G.N. Greaves, Contrib. Mineral Petrol 105 (1990) 359-368

F. Lanza and E. Parnisari, Th leaching of plutonium loaded borosilicate glass, European Appl. Res. Rept.-Nucl. Sci. Technol. 4 (1983) 1151-1170

D.J. Lam, B.W. Veal, H. Chen und G.S. Knapp, in Scientific Basis for Nuclear Waste Management Vol 1., (1979) 97-107

C. Lopez, X. Deschanels, J.M. Bart, J.M. Boubals, C. Den Auwer, E. Simoni, Solubility of actinide surrogates in nuclear glasses, Journal of Nuclear Materials 312 (2003) 76–80

B. Luckscheiter, Nesovic, M.: HLW-glass dissolution and co-precipitation studies. Mater. Res. Soc.Symp. Proc. 932, 361–368 (2006)

E. Malchukova, B. Boizot, D. Ghaleb, G. Petite, β-Irradiation effects in Gd-doped borosilicate glasses studied by EPR and Raman spectroscopies, Journal of Non-Crystalline Solids 352 (2006) 297–303

T. Maeda, T. Banba, K. Sonoda, Y. Inagaki, H. Furuya, Release and retention of uranium during glass dissolution, Journal of Nuclear Materials 298, 163-167 (2001)

O.J. McGann, P.A. Bingham, R.J. Hand, A.S. Gandy, M. Kavcic, M. Zitnik, K. Bucar, R. Edge, N.C. Hyatt, The effects of c-radiation on model vitreous wasteforms intended for the disposal of intermediate and high level radioactive wastes in the United Kingdom, Journal of Nuclear Materials, 429, (2012) 353-367

R.K. Mishra, V. Sudarsan, A.K. Tyagi, C.P. Kaushik, Kanwar Raj, S.K. Kulshreshtha, Structural studies of $ThO_2$ containing barium borosilicate glasses, Journal of Non-Crystalline Solids 352 (2006) 2952–2957

L. R. Morss, C.J. Mertz, A. J. Kropf, and J. L. Holly, Properties of Plutonium-Containing Colloids Released from Glass-Bonded Sodalite Nuclear Waste Form, Mat. Res. Soc. Symp. Proc. 713 (2002)

I. Munier, J.-L. Crovisier, B. Grambow, B. Fritz, A. Clément; "Modelling the alteration gel composition of simplified borosilicate glasses by precipitation of an ideal solid solution in equilibrium with the leachant" Journal of Nuclear Materials, Volume 324, Issues 2-3, (2004), Pages 97-115

A. Muñoz-Páez, M.D. Alba, R. Alvero, A.I. Becerro, M.A. Castro, J.M. Trillo, Nucl. Instr. Methods in Phys. Research B 97 (1995) 142-144

N. Ollier, PhD thesis, University Lyon I, 2002

N. Ollier, M. Guittet, M. Gautier-Soyer, G. Panczer, B. Champagnon, P. Jollivet, U environment in leached SON68 type glass: a coupled XPS and time-resolved photoluminescence spectroscopy study, Optical Materials 24 (2003) 63–68

D. Petit-Marie, J. Petiau, G. Calas und N. Jacquet-Francillon, J. Physique Colloque C8 Suppl. 12, 47, C8-849 (1986)

S. Peuget, J.-N. Cachia, C. Jegou, X. Deschanels, D. Roudil, V. Broudic, J.M. Delaye, J.-M. Bart, Irradiation stability of R7T7-type borosilicate glass, Journal of Nuclear Materials 354 (2006) 1–13

E.M. Pierce, B.P. McGrail, P.F. Martin, J. Marra, B.W. Arey, K.N. Geiszler Accelerated weathering of high-level and plutonium-bearing lanthanide borosilicate waste glasses under hydraulically unsaturated conditions, Applied Geochemistry 22 (2007) 1841–1859

Prochazka / Journal of Non-Crystalline Solids 353 (2007) 2052–2056

A. Quintas, O. Majerus, M. Lenoir, D. Caurant, K. Klementiev, A. Webb. Effect of alkali and alkaline-earth cations on the neodymium environment in a rare-earth rich aluminoborosilicate glass, Journal of Non-Crystalline Solids 354 (2008) 98–104

Rai, D., Yui, M., Moore, D.A.: Solubility and solubility product at 22 °C of UO2(c) precipitated from aqueous U(IV) solutions. J. Solution Chem. 32, 1–17 (2003)

Rai, D., Yui, M., Hess, N.J., Felmy, A.R., Moore, D.A.: Thorium reactions in borosilicate-glass/watersystems. Radiochim. Acta 93, 443–455 (2005)

D. Rai, M. Yui, A. Kitamura, B. Grambow, Thermodynamic Approach for Predicting Actinide and Rare Earth Concentrations in Leachates from Radioactive Waste Glasses, J Solution Chem (2011) 40:1473–1504

Ribet, I., Gin, S., Godon, N., Jollivet, P.,Minet, Y., Grambow, B., Abdelouas, A., Ferrand, K., Lemmens, K., Aertsens, M., Pirlet, V., Jacques, D., Crovisier, J.L., Aouad, G., Arth, A., Clement, A., Fritz, B., Morvan, G., Munier, I., Del Nero, M., Ozgumus, A., Curti, E., Luckscheiter, B., Schwyn, B.: Longterm behavior of glass: improving the glass source term and substantiating the basic hypotheses. Final technical report, Contract N FIKW-CT-2000-00007, European Commission (2004)

H.D. Schreiber, G.B. Balazs, Phys. Chem. Glasses 23 (5) (1982a) 139.

H.D. Schreiber, G.B. Balazs, P.L. Jamison, A.P. Shaffer, Phys. Chem. Glasses 23 (5) (1982b) 147.

A.A. Shiryaev, Y.V. Zubabichus, S.V. Stefanovsky, A.G. Ptashkin, J.C. Marra, XAFS of Pu and HF LIII Edge in Lanthanide-Borosilicate Glass, Mat. Res. Soc. Symp. Proc. 1193 (2009) 259-265

R.N. Sinclair, J.A.E. Desa und A.C. Wright, J. Am. Ceram. Soc. 66, 72 (1983)

T. Suzuki, Mécanismes d'altération des matrices de déchets HA–VL sous irradiation, PhD thesis, University of Nantes, France (2008)

P. van Iseghem, M. Aertens, S. Gin, D. Deneele, B. Grambow, D. Strachan, P. McGrail, G. Wicks, GLAMOR-Or how we achieved a common understanding on the decrease of glass dissolution kinetics. Environmental Issues and Waste Management Technologies in the Materials and Nuclear Industries XII: Ceramic Transactions, Vol 207 Alex Cozzi (Editor), Tatsuki Ohji (Editor), 2009, pp 115-126

R.W.Veal et al. in Handbook of the Physics and chemistry of the Actinides (1987) 271

E.Y.Vernaz, Mat. Res. Soc. Symp. Proc. 257, (1992) 37-48

L.Wang, P. Van Iseghem and A. Maes, Plutonium Leaching from a Reference Nuclear Waste Glass in Synthetic Interstitial Claywater, Mat. Res. Soc. Symp. Proc. 294, (1992)

W.J. Weber, R.C. Ewing, C.A. Angell, G.W. Arnold, A.N. Cormack, J.M. Delaye, D.L. Griscom, L.W. Hobbs, A. Navrotsky, D.L. Price, A.M. Stoneham, M.C. Weinberg, J. Mater. Res. 12 (1997) 1946.

D.M. Wellman, J.P. Icenhower, W.J. Weber, Elemental dissolution study of Pu-bearing borosilicate glasses, Journal of Nuclear Materials 340 (2005) 149–162

D.J. Wronkiewicz, L.M. Wang, J.M. Bates and B.S. Tani, Mat. Res. Soc. Symp. Proc. 294, (1993) 183-206

Xia, Y., Rao, L., Rai, D., Felmy, A.R.: Determining the distribution of Pu, Np, and U oxidation states in dilute NaCl and synthetic brine solutions. J. Radioanal. Nucl. Chem. 250, 27–37 (2011)

G. Yang, G. Möbus, R.J. Hand, Cerium and boron chemistry in doped borosilicate glasses examined by EELS, Micron 37 (2006) 433–441

G. Yang, S. Cook, R.J. Hand, G. Möbus, $CeO_2$ nano-precipitation in borosilicate glasses: A redox study using EELS, Journal of the European Ceramic Society 30 (2010) 831–838